\documentclass[12pt]{article}
\pdfoutput=1
\usepackage{jheppub}

\newlength{\abstractwidth}
\usepackage{amsthm}
\usepackage{amsmath}
\usepackage{amsfonts}
\usepackage{amssymb}
\usepackage{latexsym}

\usepackage{epsf}
\usepackage{color}
\usepackage{graphicx}
\usepackage{tikz}
\usepackage{dsfont}

\usetikzlibrary{arrows,shapes,positioning}
\usetikzlibrary{decorations.markings}
\usepackage[rightcaption]{sidecap}
\tikzstyle arrowstyle=[scale=1]
\tikzstyle directed=[postaction={decorate,decoration={markings,
    mark=at position .65 with {\arrow[arrowstyle]{stealth}}}}]
\tikzstyle reverse directed=[postaction={decorate,decoration={markings,
    mark=at position .65 with {\arrowreversed[arrowstyle]{stealth};}}}]
\usetikzlibrary{positioning}

    \usepackage{hyperref}
\definecolor{darkred}{rgb}{0.8,0.1,0.1}
\hypersetup{colorlinks=true, linkcolor=darkred, citecolor=blue, linktoc=page}

\renewcommand{\thanks}[1]{\footnote{#1}}

\newcommand{\bea}{\begin{eqnarray}}
\newcommand{\eea}{\end{eqnarray}}
\newcommand{\be}{\begin{eqnarray}}
\newcommand{\ee}{\end{eqnarray}}
\newcommand{\bma}{\begin{matrix}}
\newcommand{\ema}{\cr\end{matrix}}

\def\cF{{\cal F}}

\def\cN{{\cal N}}

\def\cT{{\cal T}}

\def\cZ{{\cal Z}}

\def\ZZ{{\mathbb Z}}

\def\tr{{\rm tr}}
\def\Tr{{\rm Tr}}

\def\half{{1\over 2}}
\def\p{\partial}

\def\a{\alpha}
\def\m{\mu}
\def\b{\beta}
\def\g{\gamma}

\def\l{\lambda}

\def\f{\varphi}
\def\n{\nu}

\def\no{\nonumber}

\def\Arf{{\mathrm{Arf}}}
\def\Pin{{\mathrm{Pin}}}

\def\f{{\sf f}}
\def\fl{{\sf f_L}}
\def\fr{{\sf f_R}}
\def\FL{{\sf F_L}}

\def\F{{\sf F}}

\makeatletter
\def\@fpheader{\ }
\makeatother
\title{Stable Vacua for Tachyonic Strings}
\author[1]{Justin Kaidi}
\emailAdd{jkaidi@scgp.stonybrook.edu}

\affiliation[1]{ Simons Center for Geometry and Physics\\
Stony Brook University\\
Stony Brook NY 11794, USA\\[-4mm]}

\abstract{In addition to the familiar string theories with spacetime supersymmetry, there exist a number of non-supersymmetric ten-dimensional superstrings. Nearly all of these theories have closed string tachyons, indicating a misidentification of the vacuum around which they are to be quantized. In this work, we identify (meta)stable vacua for all known tachyonic superstrings. These vacua are all lower-dimensional, with most of them being familiar two-dimensional string theories. However, there are also intriguing examples of solutions in dimensions 6, 8, and 9, with respective gauge groups $E_7 \times E_7$, $SU(16)$, and $E_8$.  These special vacua have positive vacuum energy and no moduli.
}

\begin{document}

\maketitle
\section{Introduction}

One of the long-touted virtues of string theory is its apparent uniqueness. Famously, the Type IIA/B, heterotic $SO(32)$, heterotic $E_8 \times E_8$, and Type I superstring theories have all been shown to be different perturbative descriptions of a single underlying theory. These five descriptions are typically given as expansions around ten-dimensional flat space, where they are tachyon-free and have manifest spacetime supersymmetry. 

However, it has long been known that there exist additional, non-supersymmetric superstring theories, which lie outside of the standard duality web. These include the Type 0A/B theories, eight $\rm Pin^-$ Type 0 theories \cite{Bergman:1997rf,Bergman:1999km,Kaidi:2019pzj,Kaidi:2019tyf} and a series of seven non-supersymmetric heterotic strings \cite{AlvarezGaume:1986jb,Dixon:1986iz,Kawai:1986vd}. 
Though some preliminary work has been done on bringing these theories into contact with the more familiar superstrings \cite{Itoyama:1986ei,Ginsparg:1986wr,Blum:1997cs,Blum:1997gw,Bergman:1997rf,Bergman:1999km,Suyama:2001bn,Hellerman:2007fc}, as well as exploring their connections to phenomenology \cite{Faraggi:2019fap,Faraggi:2019drl,Faraggi:2020wej}, it is safe to say that this is not yet a completed endeavor. 

One of the reasons that these additional string theories have remained relatively understudied is that almost all of them are plagued by closed string tachyons.\footnote{Notable exceptions are the $\Pin^+$ Type 0 strings \cite{Kaidi:2019tyf}, the $O(16)\times O(16)$ heterotic string \cite{AlvarezGaume:1986jb}, and the Type $\tilde{\rm I}$ string \cite{Sugimoto:1999tx}, all of which are tachyon-free.} But the presence of a tachyon is not itself indicative of an inconsistency in the theory. Instead, it simply means that the presumed vacuum (typically ten-dimensional Minkowski space) is not a legitimate solution. In other words, these additional strings theories should not be interpreted as describing fluctuations around ten-dimensional flat space, but rather as fluctuations about some other, possibly lower-dimensional background.
As long as \textit{some} stable vacuum exists, these theories are well-defined quantum gravitational theories, and are thus expected on general grounds to fit into the string duality web. In this paper, we will find such (meta)stable vacua for the tachyonic theories listed above, using techniques developed in \cite{Hellerman:2004qa,Hellerman:2006nx,Hellerman:2006ff,Hellerman:2006hf,Hellerman:2007fc,Hellerman:2007zz}. 

Seemingly unrelatedly, it has also been known that there exist a number of consistent two-dimensional string theories. These include three heterotic strings \cite{McGuigan:1991qp,Giveon:2004zz,Davis:2005qe,Seiberg:2005nk,Davis:2005qi}, two oriented Type 0 strings \cite{Takayanagi:2003sm,Douglas:2003up}, and a total of eight unoriented Type 0 strings \cite{Bergman:2003yp,Gomis:2005ce}. Though much progress has been made in understanding these theories via e.g. matrix model techniques of \cite{McGreevy:2003kb,Gomis:2003vi}, these two-dimensional theories have remained largely disconnected from the space of ten-dimensional strings. Nevertheless, since these theories describe consistent quantum gravity theories (albeit in two dimensions), by the string uniqueness principle they are again expected to be continuously connected to the space of higher-dimensional string theories. There is one situation in which such a connection has been concretely realized -- namely, in \cite{Hellerman:2007fc} it was shown that closed string tachyon condensation in the ten-dimensional oriented Type 0 theories leads to a dimension-reducing cascade, with the endpoint being the two-dimensional Type 0 theories. Many of the stable vacua studied in this paper will be exactly of this type, i.e. we will see that most of the tachyonic ten-dimensional strings admit a dynamical transition to one of the two-dimensional theories previously studied in the literature. This paves the way towards bringing these 2d strings into contact with the usual duality web.\footnote{Additional two-dimensional string theories include the 2d Type II \cite{Takayanagi:2004ge,Seiberg:2005bx,Ita:2005ne} and Type I \cite{Gomis:2005ce} theories. It is natural to suspect that these too can be obtained via dynamical transitions from tachyonic ten-dimensional strings. Indeed, for the Type II theories a potential candidate in 10d is the $(n_{even} , n_{odd}) = (6,4)$ element of the ``Seiberg series" of \cite{Hellerman:2007fc}. Since these tentative ten-dimensional starting points are not as well-known though, we will not discuss them further.}

There are however some interesting cases in which tachyonic ten-dimensional string theories do \textit{not} admit a stable 2d vacuum; at least not via the simplest condensation mechanism. For those cases, we will instead identify stable vacua in dimensions $d>2$. Concretely, for the heterotic $E_8$, $U(16)$, and $(E_7 \times SU(2))^2$ heterotic strings, we will identify stable vacua in dimensions $d=9,8,$ and 6.\footnote{The first of these has already been studied in \cite{Hellerman:2007zz}.} All of the resulting lower-dimensional theories are free of perturbative anomalies, and interestingly all have positive one-loop cosmological constant. Because these theories have no moduli (and in particular the dilaton has an effective mass), the dilaton tadpole implied by the positive cosmological constant is not expected to signal any sort of perturbative instability. Hence we suspect that these cases represent metastable solutions of string theory with positive vacuum energy, at least in the weak-coupling regime.

This paper is organized as follows. In Section \ref{sec:hetstrings} we discuss non-supersymmetric heterotic string theories. After reviewing the six tachyonic ten-dimensional theories in Section \ref{sec:10dheteroticstrings}, as well as the three two-dimensional theories in Section \ref{sec:2dhetstring}, we proceed to a discussion of tachyon condensation. In Section \ref{sec:condensation} we consider condensation to theories in $d>2$, study local and global anomaly cancellation in the resulting theories, and compute the cosmological constant. In Section \ref{sec:hettach2d}, we consider condensation of the remaining heterotic theories to $d=2$. In Section \ref{sec:N11strings} we consider tachyon condensation of oriented and unoriented Type 0 string theories, where we will see that all cases can be condensed to known two-dimensional strings.

\section{Heterotic strings}
\label{sec:hetstrings}
\subsection{Heterotic strings in $d=10$}
\label{sec:10dheteroticstrings}
We begin by reviewing the non-supersymmetric heterotic string theories obtained in \cite{Dixon:1986iz,Kawai:1986vd}. Our construction will be somewhat different from that of the original works.

 First recall that in the free fermion formulation, the worldsheet theory of heterotic strings consists of ten bosonic fields $X^\m$, ten right-moving fermionic fields $\psi^\m$, and thirty-two left-moving fermionic fields $\tilde \lambda^a$. The supersymmetric heterotic strings are taken to have independent spin structures for the left- and right-moving fermions, both of which are summed over in the GSO projection. In contrast, non-supersymmetric heterotic strings are obtained by identifying the left- and right-moving spin structures. The simplest non-supersymmetric heterotic string has torus partition function $Z=Z_{\rm NS} - Z_{\rm R}$, where
\bea
\label{eq:SO32case}
Z_{\rm NS} &=& \half |\eta|^{-16} \left[(Z^0_0)^8 (\overline{Z^0_0})^{32} -(Z^0_1)^8 (\overline{Z^0_1})^{32}  \right]~,
\no\\
Z_{\rm R} &=& \half |\eta|^{-16} (Z^1_0)^8 (\overline{Z^1_0})^{32} ~.
\eea
The minus sign in the NS-sector partition function comes from the fact that the superghost ground state is fermionic, and the overall factor of $1/2$ is added for the sum over spin structure. We use the usual notation that 
\bea
\label{eq:Zabdef}
Z^\a_\b = {\Bigg(}{\vartheta_{\a \b} (0| \tau)\over \eta(\tau)}{\Bigg)}^\half~,
\eea
with $\vartheta_{\a \b}(z|\tau)$ the Jacobi theta functions and $\eta(\tau)$ the Dedekind eta function. 

The spectrum of the theory in (\ref{eq:SO32case}) can be easily read off by considering the level-matched partition functions $\cZ_{\rm NS,R}$, obtained by integrating over $\tau_1$. This gives
\bea
\cZ_{\rm NS} &=& 32 \,(q \bar q)^{-\half} + 4032 + 188928 \,(q \bar q)^\half + \dots ~,
\no\\
\cZ_{\rm R} &=& 8388608\, q \bar q + 3019898880\,(q \bar q)^2 + \dots~.
\eea
From the first line, we see that there are 32 tachyons, as well as 4032 massless bosonic degrees of freedom. These correspond to a graviton (35), B-field (28), dilaton (1), and a remaining 496 gauge bosons of $SO(32)$. There are no massless fermions, so it is clear that this theory is not supersymmetric.

We may obtain other non-supersymmetric strings from this $SO(32)$ theory as follows. We begin by noting that the system of 32 left-moving fermions admits a $(\ZZ_2)^5$ global symmetry, with generators
\bea
g_1 &=& \sigma_3 \otimes \mathds{1}_2  \otimes \mathds{1}_2  \otimes \mathds{1}_2  \otimes \mathds{1}_2~,  \hspace{0.8 in}g_2 =  \mathds{1}_2  \otimes \sigma_3\otimes \mathds{1}_2  \otimes \mathds{1}_2  \otimes \mathds{1}_2 ~,
\no\\
g_3 &=& \mathds{1}_2  \otimes \mathds{1}_2  \otimes \sigma_3 \otimes \mathds{1}_2  \otimes \mathds{1}_2~,  \hspace{0.8 in}g_4 =  \mathds{1}_2  \otimes  \mathds{1}_2  \otimes \mathds{1}_2  \otimes\sigma_3\otimes \mathds{1}_2 ~,
\no\\
g_5 &=& \mathds{1}_2  \otimes \mathds{1}_2  \otimes \mathds{1}_2  \otimes \mathds{1}_2 \otimes \sigma_3~.
\eea
Each of these acts on 16 of the 32 left-moving fermions $\tilde \lambda^a$ with a minus sign, and acts as the identity on the others. We may now choose to gauge some or all of this global symmetry. Gauging $(\ZZ_2)^n$ for $0\leq n \leq 5$ means that we should replace the above partition function with
\bea
\label{eq:gaugedZs}
Z_{\rm NS} &=& \half |\eta|^{-16} \left[(Z^0_0)^8 L^0_0 -(Z^0_1)^8 L^0_1 \right]~,
\no\\
Z_{\rm R} &=& \half |\eta|^{-16} (Z^1_0)^8 L^1_0~,
\eea
where $L^a_b$ are twisted left-moving partition functions,\footnote{One could contemplate obtaining new theories by allowing for discrete torsion in these sums. That is, we allow for some extra phases dictated by an element of $H^2((\ZZ_2)^{n-1}, U(1))$. This is analogous to what was done for the supersymmetric $E_8 \times E_8$ string to obtain the non-tachyonic $O(16) \times O(16)$ heterotic string. We could in fact go further and modify the twisted partition functions by a general element of $\mho^2_{\rm Spin}(B(\ZZ_2)^{n-1}):= {\rm Hom}(\Omega_2^{\rm Spin}(B(\ZZ_2)^{n-1}),U(1))$, much in the spirit of \cite{Kaidi:2019pzj,Kaidi:2019tyf}. We save this analysis for a separate work.}
\bea
\label{eq:gaugedLs}
L^0_0 &=& {1 \over 2^n}(\overline{Z^0_0})^{16}\left[(\overline{Z^0_0})^{16} + (2^n -1)(\overline{Z^0_1})^{16} + (2^n -1)(\overline{Z^1_0})^{16} \right]~,
\no\\
L^0_1 &=& {1 \over 2^n}(\overline{Z^0_1})^{16}\left[(\overline{Z^0_1})^{16} + (2^n -1)(\overline{Z^0_0})^{16} + (2^n -1)(\overline{Z^1_0})^{16} \right]~,
\no\\
L^1_0 &=& {1 \over 2^n}(\overline{Z^1_0})^{16}\left[(\overline{Z^1_0})^{16} + (2^n -1)(\overline{Z^0_0})^{16} + (2^n -1)(\overline{Z^0_1})^{16} \right]~.
\eea
It is easy to verify that the full partition function, i.e. $Z= Z_{\rm NS}-Z_{\rm R}$, is modular invariant for all $n$.

With these expressions, we may now read off the spectra by considering the level-matched partition functions, 
\bea
n&=&1: \hspace{0.2 in} \cZ_{\rm NS}= 16 \,(q \bar q)^{- \half} + 3008 + 168192 \,(q \bar q)^\half + \dots
\no\\
&\vphantom{.}& \hspace{0.51 in} \cZ_{\rm R} = 2048 + 8912896\, q \bar q + \dots 
\no\\\no\\
n&=&2: \hspace{0.2 in} \cZ_{\rm NS}= 8 \,(q \bar q)^{- \half} + 2496 + 157824 \,(q \bar q)^\half + \dots
\no\\
&\vphantom{.}& \hspace{0.51 in} \cZ_{\rm R} = 3072 + 9175040\, q \bar q + \dots 
\no\\\no\\
n&=&3: \hspace{0.2 in} \cZ_{\rm NS}= 4 \,(q \bar q)^{- \half} + 2240 + 152640 \,(q \bar q)^\half + \dots
\no\\
&\vphantom{.}& \hspace{0.51 in} \cZ_{\rm R} = 3584 + 9306112\, q \bar q + \dots
\no\\\no\\
n&=&4: \hspace{0.2 in} \cZ_{\rm NS}= 2 \,(q \bar q)^{- \half} + 2112 + 150048 \,(q \bar q)^\half + \dots
\no\\
&\vphantom{.}& \hspace{0.51 in} \cZ_{\rm R} = 3840 + 9371648\, q \bar q + \dots  
\no\\\no\\
n&=&5: \hspace{0.2 in} \cZ_{\rm NS}=  (q \bar q)^{- \half} + 2048 + 148752 \,(q \bar q)^\half + \dots
\no\\
&\vphantom{.}& \hspace{0.51 in} \cZ_{\rm R} = 3968 + 9404416\, q \bar q + \dots 
 \eea
  This gives us the data in Table \ref{tab:tachhet}. These theories are, respectively, the $O(16) \times E_8$, $O(8) \times O(24)$, $(E_7 \times SU(2))^2$, $U(16)$, and $E_8$ non-supersymmetric heterotic string theories.\footnote{Note that we could also obtain most of these theories via the covariant lattice formulation of \cite{Lerche:1986he,Lerche:1986ae,Lust:1989tj}. We do not use that construction here though since it does not give the $E_8$ theory. This is because in that case the fermions cannot all be bosonized to lattice bosons.} 
Let us note that we are not being particularly careful about the global structure of the spacetime gauge group here. The cautious reader should take all of our statements at the level of the  algebra.
 \begin{table}[!htbp]
\begin{center}
\begin{tabular}{c|c|c|c}
$n$ & tachyons & massless fermions & gauge bosons
\\\hline
0 & 32 & 0 & 496
\\
1 & 16 & 256 & 368 
\\
2 & 8 & 384 & 304 
\\ 
3 & 4 & 448 & 272 
\\
4& 2 & 480 & 256 
\\
5 & 1 & 496 & 248
\end{tabular}
\end{center}
\caption{Tachyonic heterotic strings in $d=10$.}
\label{tab:tachhet}
\end{table}%

An intuitive explanation for the first four of the above gauge groups is as follows. As said before, each of the generators $g_i$ of $(\ZZ_2)^5$ acts as $-1$ on sixteen fermions and as the identity on the remaining fermions. Under any given subgroup $(\ZZ_2)^n$, $2^{5-n}$ fermions remain invariant while $2^5 (1 - {1 \over 2^n})$ change sign. We thus expect that in general one breaks $\mathfrak{so}(32)$ to $\mathfrak{so}(2^{5-n}) \times \mathfrak{so}(2^5 (1 - {1 \over 2^n}))$. This leads to an expected sequence of gauge algebras $\mathfrak{so}(16) \times \mathfrak{so}(16)$, $\mathfrak{so}(8) \times \mathfrak{so}(24)$, $\mathfrak{so}(4) \times \mathfrak{so}(28)$, and $\mathfrak{so}(2) \times \mathfrak{so}(30)$. Indeed, the $\mathfrak{so}(2^{5-n})$ factor is always present in the final answer (recall $\mathfrak{su}(2) \times \mathfrak{su}(2) \cong \mathfrak{so}(4)$ and $\mathfrak{u}(16) \cong \mathfrak{so}(2) \times \mathfrak{su}(16)$), but it turns out that the $\mathfrak{so}(2^5 (1 - {1 \over 2^n}))$ factor generically gets enhanced to a larger algebra of the same rank. A crucial point is that in all cases, the tachyon is a vector of $\mathfrak{so}(2^{5-n})$, hence its contribution of $2^{5-n}$ to the partition function. We will discuss this further in Section \ref{sec:condensation}. 

It will be useful to have more detail on the chiral matter content of each theory. In self-evident notation, we find the following massless fermions, 
\begin{align}
\label{eq:chiralfermdetails}
&SO(32):  \hspace{1.1 in}\mathrm{none}
\no\\
&O(16)\times E_8: \hspace{0.8 in}(\mathbf{128},1)_+~, \,\,\,\, \,(\mathbf{128'},1)_-
\no\\
&O(24) \times O(8): \hspace{0.65 in} (\mathbf{24},\mathbf{8_v})_+~, \,\,\,\,\, (\mathbf{24}, \mathbf{8_c})_-
\no\\
&(E_7 \times SU(2))^2:\hspace{0.58 in} (\mathbf{56},\mathbf{2};1,1)_+\oplus(1,1;\mathbf{56},\mathbf{2})_+~, \,\,\,\,\,(\mathbf{56},1;1,\mathbf{2})_-\oplus(1,\mathbf{2};\mathbf{56},1)_-
\no\\
&SU(16) \times U(1): \hspace{0.55 in}  (\mathbf{120},2)_+\oplus(\mathbf{\overline{120}},-2)_+~, \,\,\,\,\, (\mathbf{120},-2)_-\oplus(\mathbf{\overline{120}},2)_-
\no\\
&E_8: \hspace{1.45in} \mathbf{248}_+~, \,\,\,\,\, \mathbf{248}_-
\end{align}
where the subscript $\pm$ represents $\mathbf{8}_{s,c}$ of the spacetime $\mathrm{Spin}(8)$. Adding the fermions in each line gives the cumulative results of Table \ref{tab:tachhet}.

Perturbative anomalies are cancelled as follows. The $SO(32)$ and $E_8$ theories have non-chiral spectra, so these theories are trivially non-anomalous. For the $O(16)\times E_8$ and $O(24) \times O(8)$ theories, anomaly cancellation follows simply from the relations
\bea
\Tr_{\mathbf{8_v}} F^{2n} = \Tr_{\mathbf{8_s}} F^{2n} = \Tr_{\mathbf{8_c}} F^{2n}~, \hspace{0.5 in} \Tr_{\mathbf{128}} F^{2n} = \Tr_{\mathbf{128'}} F^{2n}~.
\eea
Finally for the $(E_7 \times SU(2))^2$ and $U(16)$ theories, anomalies can be cancelled by addition of appropriate Green-Schwarz terms.

\subsection{Heterotic strings in $d=2$}
\label{sec:2dhetstring}
In addition to the non-supersymmetric heterotic strings reviewed above, there are three heterotic strings in two dimensions. Two of these strings were introduced in \cite{McGuigan:1991qp,Giveon:2004zz,Davis:2005qe}, while the remaining one was discussed in \cite{Seiberg:2005nk,Davis:2005qi}.

In order to ensure worldsheet anomaly cancellation in two dimensions, we consider solutions with a linear dilaton. In particular, we take $\phi \propto X_1$ with the proportionality constant chosen such that the system has central charge $c_{X_1}=13$. The right-moving theory is composed of the fields $X_{0,1}$ together with superpartners $\psi_{0,1}$, giving a total central charge $c_R = 15$. The left-moving theory is a bosonic string with $X_1, X_0$, and an additional twelve bosons, for a total central charge of $c_L = 26$. The twelve extra left-moving bosons must be compactified on an even self-dual lattice to give a sensible two-dimensional spacetime interpretation. The candidate lattices are the root lattices of $O(8) \times E_8$ and $O(24)$.

The two-dimensional heterotic strings may be given a free fermion interpretation as follows. We begin by replacing the twelve left-moving bosons with 24 free fermions. Then as in ten dimensions, different theories correspond to different gaugings of discrete symmetries of the system of fermions. As before, we may begin with the ungauged case, which in other words corresponds to assigning all fermions a single spin structure. This gives rise to the partition function 
\bea
\label{eq:2doriginalhet}
Z_{\rm NS} = \half \left[(\overline{Z^0_0})^{24} - (\overline{Z^0_1})^{24} \right] ~,\hspace{0.8 in} Z_{\rm R} = \half ( \overline{Z^1_0})^{24} ~,
\eea
where the sign in the NS partition function comes from right-moving superghosts. It is a convenient modular accident that $Z=Z_{\rm NS}-Z_{\rm R} = 24$, which is trivially modular invariant. The level-matched partition functions are 
\bea
\label{eq:2dO24cZ}
\cZ_{\rm NS} = 24~, \hspace{0.8 in} \cZ_{\rm R} = 0~.
\eea
The spacetime content of this theory is thus a set of 24 massless bosons transforming in the fundamental of $O(24)$, with the gravitons and $O(24)$ gauge bosons not contributinig any propagating degrees of freedom in two dimensions. There are however discrete graviton and gauge fields, where by discrete here we mean that the corresponding vertex operators exist at only discrete values of momenta (namely $p=0$) \cite{Davis:2005qe}. 

This worldsheet theory enjoys a $(\ZZ_2)^2$ symmetry, with each element of the group acting on 16 of the left-moving fermions with a sign, and leaving the remaining 8 invariant. Gauging $(\ZZ_2)^n$ for $0 \leq n \leq 2$ then gives rise to the following partition functions, 
\bea
\label{eq:2dnhet1}
Z_{\rm NS} = \half \left( L^0_0 - L^0_1 \right)~, \hspace{0.5 in}Z_{\rm R} = \half L^1_0~,
\eea
where $L^a_b$ are now defined as
\bea
\label{eq:2dnhet2}
L^0_0 &=& {1 \over 2^n}(\overline{Z^0_0})^{8}\left[(\overline{Z^0_0})^{16} + (2^n -1)(\overline{Z^0_1})^{16} + (2^n -1)(\overline{Z^1_0})^{16} \right]~,
\no\\
L^0_1 &=& {1 \over 2^n}(\overline{Z^0_1})^{8}\left[(\overline{Z^0_1})^{16} + (2^n -1)(\overline{Z^0_0})^{16} + (2^n -1)(\overline{Z^1_0})^{16} \right]~,
\no\\
L^1_0 &=& {1 \over 2^n}(\overline{Z^1_0})^{8}\left[(\overline{Z^1_0})^{16} + (2^n -1)(\overline{Z^0_0})^{16} + (2^n -1)(\overline{Z^0_1})^{16} \right]~.
\eea
As in (\ref{eq:2doriginalhet}), there are no right-moving contributions to the partition function, since in 2d they are completely cancelled by ghosts.

An alternative interpretation of these partition functions is as follows. First, the $n=1$ gauging can be interpreted as starting with the $O(24)$ theory, splitting the left-moving fermions into groups of $8$ and $16$, and giving them separate spin structures. This naively breaks $O(24)$ to $O(8) \times O(16)$, but the $O(16)$ experiences an enhancement to $E_8$, as evident from the fact that the partition function contains a factor of the $E_8$ characters,
\bea
Z_{\rm NS} = {1 \over 2} \left( (\overline{Z^0_0})^8 - (\overline{Z^0_1})^8\right)\chi_{E_8}(\bar q)~, \hspace{0.5 in}Z_{\rm R} = {1 \over 2}\, (\overline{Z^1_0})^8\,\chi_{E_8}(\bar q)~.
\eea
Hence this theory has $O(8) \times E_8$ gauge group. 

 By the abstruse identity, in this case the full partition function vanishes, i.e. $Z = Z_{\rm NS} - Z_{\rm R}=0$, and is thus trivially modular invariant. The vanishing of the partition function implies that the spectrum of the theory is supersymmetric. However, the linear dilaton spoils spacetime supersymmetry. The level-matched partition functions are concretely
\bea
\label{eq:2dO8E8cZ}
n=1: \hspace{0.4 in}\cZ_{\rm NS} = 8~, \hspace{0.3 in} \cZ_{\rm R} = 8~.
\eea
This corresponds to eight massless bosons in the $\mathbf{8}_v$ of $O(8)$, eight right-moving fermions in the $\mathbf{8}_s$, and eight left-moving fermions in the $\mathbf{8}_c$. The corresponding vertex operators are known \cite{Davis:2005qe}, though we will not need them here.

As for the $n=2$ gauging, it can be interpreted as gauging left-moving worldsheet fermion number $(-1)^\fl$ in the $O(24)$ theory. At the level of the partition function, this can be seen as follows. Since the right-movers do not contribute to the partition function, the gauging of this symmetry changes the 24 massless scalars of the $O(24)$ theory to $24$ chiral fermions. Hence the partition function is $- \half$ times the original $O(24)$ partition function, i.e. $Z= - \half (24) = -12$. It can be checked that this matches the result obtained from (\ref{eq:2dnhet1}) and (\ref{eq:2dnhet2}) with $n=2$.

From this result it follows immediately that the level-matched partition functions are 
\bea
\label{eq:2dO8O24cZ}
n=2: \hspace{0.4 in}\cZ_{\rm NS} = 0~, \hspace{0.3 in} \cZ_{\rm R} = 12~.
\eea
The $O(24)$ gauge symmetry remains unbroken, and we see that we have $24$ chiral fermions transforming in the fundamental. This matter content is seemingly anomalous, but as discussed in \cite{Seiberg:2005nk} the anomaly is cancelled by a two-dimensional analog of the Green-Schwarz mechanism. We will return to this point in Section \ref{sec:hettach2d}.

\subsection{Tachyon condensation to $d=6,8,9$}
\label{sec:condensation}
We now consider condensation of the tachyon in the ten-dimensional non-supersymmetric heterotic string theories, following the general strategy of \cite{Hellerman:2007zz}. If we denote the left-moving fermions transforming in the vector of $SO(2^{5-n})$ by $\tilde \lambda^a$ for $a = 1, \dots, 2^{5-n}$, then the tachyon appears in the worldsheet action via a superpotential term 
\bea
W = \sum_{a=1}^{2^{5-n}} \tilde \lambda^a\, \cT^a(X)~.
\eea
This superpotential induces a scalar potential which takes the form
\bea
\label{eq:hetscalarpot}
V = {1 \over 8 \pi } \sum_a \cT^a(X)^2  -{i \over 2 \pi} \sqrt{\a' \over 2}\sum_a \p_\m \cT^a(X)\,\tilde{\l}^a \psi^\m~.
\eea
The linearized equation of motion for the tachyon is found to be 
\bea
\label{eq:tacheom}
\p^\m\p_\m \cT^a - 2 v^\m \p_\m \cT^a + {2 \over \a'} \cT^a = 0 
\eea
where $v^\m = \p^\m \phi$ is the dilaton gradient. For reasons to be seen, we will choose a lightlike linear dilaton profile of the form 
\bea
\label{eq:higherdimdil}
\phi =- {\g \over \sqrt{2 \a'}} X^-~,
\eea
where $X^\pm = {1 \over \sqrt{2}}(X^0 \pm X^1)$ and $\g>0$ is a constant to be fixed momentarily.

 There are certain solutions to (\ref{eq:tacheom}) which give rise to tachyon superpotentials that are exactly marginal. One such solution is to consider a lightlike tachyon profile \cite{Hellerman:2004qa,Hellerman:2006nx,Hellerman:2006ff,Hellerman:2006hf}
\bea
\label{eq:tachprof}
\cT^a(X) = \sqrt{2 \over \a'} \,e^{\b X_+} \sum_{i=2}^{r+1} M^a_i X^i~,\hspace{0.3 in} \b : =\sqrt{2 \over \a'\g^2}~,
\eea
where $M^a_i$ is a matrix chosen such that $\cT^1 = \dots = \cT^{r} = 0$ along some unique locus $X^2 = \dots =X^{r+1} = 0$. In particular, as long as $r \leq 8$ we can simply take 
\bea
\label{eq:Mansatz}
M^a_i = m \,\delta^a_{i-1}~.
\eea
 Introducing such a tachyon gives rise to the scalar potential
\bea
V = {m^2 \over 4 \pi \a'}e^{ 2\b X^+} \sum_{i=2}^{r+1} (X^i)^2 -{i m \over 2 \pi} e^{ \b X^+}\sum_{a=1}^{r}\tilde{\lambda}^a (\psi^{a+1} + \b X^{a+1} \psi^+)~,
\eea
which gives rise to an exponentially-growing mass $m(X^+) := m \,e^{ \b X^+}$ for $X^i$ and $\psi^i$ $(i= 2, \dots, r+1)$, as well as $\tilde \lambda^a$ $(a= 1, \dots, r)$. Thus as $X^+ \rightarrow \infty$, fluctuations in these directions are highly suppressed and we expect to obtain a theory in spacetime dimension $d= 10 - r$. 

In the case of the heterotic theories being studied here, we would ideally like to condense all $2^{5-n}$ components of the tachyon. Clearly this will only be possible using the ansatz (\ref{eq:Mansatz}) if $n\geq2$; otherwise we would naively obtain a theory in $d=10- 2^{5-n}<0$ dimensions. Accepting this restriction for the moment, we may now fix the parameter $\g$ in terms of $n$ as follows. As $X^+ \rightarrow \infty$, we may integrate out the fields $X^i$ and $\psi^i$ $(i= 2, \dots, 2^{5-n}+1)$, as well as $\tilde \lambda^a$ $(a= 1, \dots, 2^{5-n})$. This leads to quantum corrections to the dilaton and metric, and as it turns out these corrections appear only at one-loop. The corrections can be computed exactly using the methods in \cite{Hellerman:2004qa,Hellerman:2006ff}, and are found to be 
\bea
\label{eq:oneloopcorrs}
\Delta \phi = \sqrt{2 \over \a'}{2^{3-n} \over  \g} X^+~, \hspace{0.5 in} \Delta G_{++} = - \Delta G^{--}={2^{4-n} \over \g^2}~.
\eea
In particular, we see that we have obtained a linear shift of the dilaton. This is consistent with worldsheet anomaly cancellation -- the contributions to the worldsheet central charge lost by integrating out fields must be made up for by a linear dilaton. In particular, the total central charge lost upon integrating out $X^i, \psi^i,$ and $\tilde \lambda^a$ is ${3\over 2}\times 2^{5-n}$, and hence we require
\bea
c_\phi = 6 \a' G_{\m \n} v_\m v_\n = {3}\times2^{4-n}~.
\eea
Using the results in (\ref{eq:oneloopcorrs}), this fixes 
\bea
\g = 2^{2-{n\over 2}}~.
\eea
Note in particular that for this value of $\g$, the lower-dimensional dilaton (i.e. that obtained after integrating out the massive fields) simplifies as follows, 
\bea
\label{eq:dilspacelike}
\phi =  - {\g \over \sqrt{2 \a'}} X^- +  \sqrt{2 \over \a'}{2^{3-n} \over  \g} X^+ = { 2^{2-{n\over 2}}\over \sqrt{\a'}} X^1~.
\eea
 In other words the dilaton gradient becomes \textit{spacelike}. This will be an important point in what follows. Before moving on, let us emphasize that (\ref{eq:higherdimdil}) and  (\ref{eq:tachprof}) constitute \textit{$\a'$-exact} solutions to string theory. 
 
Having constructed the relevant dimension-changing solutions, we now study the lower-dimensional theories obtained as $X^+ \rightarrow \infty$.
The partition functions for these lower-dimensional theories are obtained by simply integrating out the fields $X^i$, $\psi^i$ for $i= 2, \dots, 2^{5-n}+1$ and $\tilde \lambda^a$ for $a= 1, \dots, 2^{5-n}$. Doing so, we straightforwardly obtain
\bea
Z_{\rm NS} &=& \half |\eta|^{-16+2^{6-n}} \left[(Z^0_0)^{8-2^{5-n}} L^0_0 -(Z^0_1)^{8-2^{5-n}} L^0_1 \right]~,
\no\\
Z_{\rm R} &=& \half |\eta|^{-16+2^{6-n}} (Z^1_0)^{8-2^{5-n}} L^1_0~,
\eea
where $L^a_b$ are now given by
\bea
L^0_0 &=& {1 \over 2^n}(\overline{Z^0_0})^{16-2^{5-n}}\left[(\overline{Z^0_0})^{16} + (2^n -1)(\overline{Z^0_1})^{16} + (2^n -1)(\overline{Z^1_0})^{16} \right]~,
\no\\
L^0_1 &=& {1 \over 2^n}(\overline{Z^0_1})^{16-2^{5-n}}\left[(\overline{Z^0_1})^{16} + (2^n -1)(\overline{Z^0_0})^{16} + (2^n -1)(\overline{Z^1_0})^{16} \right]~,
\no\\
L^1_0 &=& {1 \over 2^n}(\overline{Z^1_0})^{16-2^{5-n}}\left[(\overline{Z^1_0})^{16} + (2^n -1)(\overline{Z^0_0})^{16} + (2^n -1)(\overline{Z^0_1})^{16} \right]~,
\eea
c.f. (\ref{eq:gaugedZs}) and (\ref{eq:gaugedLs}). The case of $n=5$ was studied originally in \cite{Hellerman:2007zz}.

 Integrating over $\tau_1$ gives the following level-matched results, 
\bea
\label{eq:newhetlevelmatch}
n&=&3: \hspace{0.2 in} \cZ_{\rm NS}= 1080 \,(q \bar q)^{1 \over 4} + 31360 \,(q \bar q)^{3 \over 4} \dots
\no\\
&\vphantom{.}& \hspace{0.49 in} \cZ_{\rm R} = 224 + 276480\, q \bar q + \dots
\no\\\no\\
n&=&4: \hspace{0.2 in} \cZ_{\rm NS}= 1566\, (q \bar q)^{1 \over 8} + 76440\, (q \bar q)^{5 \over 8} \dots
\no\\
&\vphantom{.}& \hspace{0.49 in} \cZ_{\rm R} = 960 + 1711104\, q \bar q + \dots  
\no\\\no\\
n&=&5: \hspace{0.2 in} \cZ_{\rm NS}=  1785 (q \bar q)^{1 \over 16}+ 108500\, (q \bar q)^{9 \over 16}\dots
\no\\
&\vphantom{.}& \hspace{0.49 in} \cZ_{\rm R} = 1984 + 4058880\, q \bar q + \dots 
 \eea
from which we read off the data in Table \ref{tab:tab1}. Note that the methods discussed here also apply to the case of $n=2$, but since this results in a two-dimensional vacuum we will wait until the next subsection to discuss it.
 \begin{table}[htp]
\begin{center}
\begin{tabular}{c|c|c|c}
$n$ & $d$ & massless fermions & gauge bosons
\\\hline
3 & 6 & 112 & 266 
\\
4& 8 & 240 & 255 
\\
5 & 9 & 248 & 248
\end{tabular}
\end{center}
\caption{Heterotic vacua in $d>2$.}
\label{tab:tab1}
\end{table}%

In more detail, the spectrum of each theory is as follows. For $n=3$, we have a 6d tachyon-free theory with gauge group $E_7 \times E_7$ and with massless fermions in the $(\mathbf{56},1)_+$ and $(1,\mathbf{56})_-$. For $n=4$, we have an 8d tachyon-free theory with gauge group $SU(16)$ and with massless fermions in the $\mathbf{120}_+$ and $\mathbf{\overline{120}}_-$.  Finally, for $n=5$ we have a 9d tachyon-free theory with gauge group $E_8$ and with a single massless fermion in the adjoint. It is easy to see how this chiral matter descends from that in (\ref{eq:chiralfermdetails}). All theories also have gravitons, $B$-fields, and dilatons for the appropriate number of spacetime dimensions. These bosonic fields are massless in the sense that there are no mass terms in the low-energy effective action, but due to the presence of the linear dilaton background the effective mass operator gets shifted \cite{chamseddine1992study} (i.e. there is non-zero vacuum energy), leading to the effective masses observed in (\ref{eq:newhetlevelmatch}). Note finally that in 6d, the $B$-field is non-chiral.

\subsubsection{Anomalies}
\label{sec:anomalies}
Let us now discuss the anomalies of the putative $d=6,8,$ and $9$ string theories identified above. The 9d theory is clearly not subject to any perturbative anomalies, so we focus on the cases of $d=6,8$ for the moment. In both cases,  anomalies can be cancelled via appropriate Green-Schwarz counterterms, which descend from their ten-dimensional counterparts.

We begin with the 6d $E_7 \times E_7$ theory. Recall that the chiral content of this theory consisted of a left-moving fermion in the $(\mathbf{56},1)$, as well as a right-moving fermion in the $(1,\mathbf{56})$. Because there are equal numbers of left- and right-movers, pure gravitational anomalies cancel automatically. The remaining anomaly 8-form is given by 
\bea
I_{(8)} \propto {1 \over 24}\left( \Tr_{(\mathbf{56},1)} F^4 -  \Tr_{(1,\mathbf{56})} F^4\right) - {1 \over 24} \left( \Tr_{(\mathbf{56},1)} F^2 -  \Tr_{(1,\mathbf{56})} F^2\right) \tr R^2 ~,
\eea
with $F$ the curvature of the $E_7 \times E_7$ bundle. We may split $F = F_1 \oplus F_2$ in terms of curvatures for the individual $E_7$ factors. As is well known, exceptional groups have no order-four Casimir invariants, and hence the terms of type $ \tr F_i^4:= \Tr_{\mathbf{56}} F_i^4$ can be factorized.  Indeed, one computes from e.g. the appendices of \cite{Erler:1993zy,Ohmori:2014kda} that 
\bea
\tr F_i^4 = {1 \over 24} \left(\tr F_i^2\right)^2~, \hspace{0.5 in} i =1,2
\eea
which allows us to rewrite the anomaly polynomial in the form 
\bea
I_{(8)} \propto \left(\half \tr R^2 - {1 \over 24} \tr F_1^2 \right)^2-\left(\half \tr R^2 - {1 \over 24} \tr F_2^2 \right)^2  ~.
\eea
Since the anomaly polynomial does not factorize completely, the anomaly cannot be cancelled by a Green-Schwarz term involving a single two-form field. Fortunately, in the current theory  the $B$-field is non-chiral -- in other words, it can be split into self-dual and anti-self-dual pieces $B_\pm$, which can then be used to write two separate Green-Schwarz terms, {\`a} la \cite{Sagnotti:1992qw}. In particular, if we take
\bea
dH_+ =  \tr R^2 - {1 \over 12} \tr F_1^2~, \hspace{0.5 in} dH_- = \tr R^2 - {1 \over 12} \tr F_2^2~,
\eea
for $H_\pm = d B_\pm$, the relevant Green-Schwarz terms are simply 
\bea
S_{\rm GS} \propto \int B_+ \wedge \left( \tr R^2 - {1 \over 12} \tr F_1^2 \right) -\int B_- \wedge \left( \tr R^2 - {1 \over 12} \tr F_2^2 \right) ~.
\eea
Note that under spacetime parity transformation we effectively interchange the two factors of $E_7$. This is consistent with the fact that the chiral spinors charged under each $E_7$ are also interchanged. Incidentally, the opposite signs between the two Green-Schwarz terms confirms that the $B$-field must non-chiral, as opposed to having two $B$-fields of the same chirality.  

Now consider the 8d $SU(16)$ theory. The chiral content of this theory is a left-moving fermion in the $\mathbf{120}$ and a right-moving fermion in the $\overline{\mathbf{120}}$. In 8d there are no (perturbative) pure gravitational anomalies, and the anomaly 10-form is 
\bea
I_{(10)} \propto {1 \over 120} \left(\Tr_{\mathbf{120}} F^5 -\Tr_{{\mathbf{\overline{120}}}} F^5   \right) - {1 \over 72}\left(\Tr_{\mathbf{120}} F^3 -\Tr_{\mathbf{\overline{{120}}}} F^3   \right) \tr R^2~,
\eea
with $F$ the $SU(16)$ curvature. In the current notation $F$ is anti-Hermitian, so we have 
\bea
\Tr_{\mathbf{120}} F^{2n+1} = - \Tr_{\mathbf{\overline{120}}} F^{2n+1} ~.
\eea
In order to use the Green-Schwarz mechanism, we must thus check that $\Tr_{\mathbf{120}} F^5$ factorizes.

To see that it does, we begin by re-expressing the trace in the anti-symmetric 2-tensor representation in terms of the trace in the fundamental. For general $SU(N)$, one finds
\bea
\Tr_{[2]} e^{i F} = \half \left(\tr e^{i F}\right)^2 -  \half \tr e^{2i F}~,
\eea
which in particular gives 
\bea
\Tr_{[2]} F^3 &=& (N-4) \tr F^3 ~,
\no\\
\Tr_{[2]} F^5 &=& (N-16) \tr F^5 + 10\, \tr F^2\tr F^3~.
\eea
Remarkably, we see that for $SU(16)$ the $\tr F^5$ term is absent from the last equation, and hence $\Tr_{\mathbf{120}}F^5$ does indeed factorize! We thus have 
\bea
I_{(10)} &\propto& {1 \over 60} \Tr_{\mathbf{120}} F^5 - {1 \over 36}\Tr_{\mathbf{120}} F^3 \, \tr R^2
\no\\
&=&{1 \over 3} \tr F^3 \left(\half \tr F^2 - \tr R^2\right)~.
\eea
Then modifying the $B$-field such that  
\bea
dH = \tr R^2 - \half \tr F^2~,
\eea
we conclude that the appropriate Green-Schwarz term is 
\bea
S_{\rm GS} \propto \int B \wedge \tr F^3 ~.
\eea
This completes the perturbative anomaly analysis. 

We now briefly discuss global anomalies. Beginning with the case of $d=6$, the full set of possible global anomalies is captured by 
\bea
\mho^7_{\rm Spin}(BE_7 \times BE_7) := \mathrm{Hom}\left(\Omega_7^{\rm Spin }(BE_7 \times BE_7), U(1) \right)~.
\eea
 For our current purposes, it suffices to treat $BE_7$ as the Eilenberg-MacLane space $K(\ZZ,4)$, in which case we can use the results of \cite{edwards1991spin} to conclude that 
\bea
\mho^7_{\rm Spin}(BE_7 \times BE_7) = 0~.
\eea
Hence this theory is free of global anomalies.

Things are somewhat less straightforward in the remaining cases. For the $d=8$ theory, the relevant group is  $\mho^9_{\rm Spin}(BSU(16)) \cong \mho^9_{\rm Spin}(pt) \oplus \tilde\mho^9_{\rm Spin}(BSU(16))$. Using the results of e.g.  Appendix B of \cite{Lee:2020ojw}, the reduced dual bordism group is seen to vanish,
\bea
 \tilde\mho^9_{\rm Spin}(BSU(16)) = 0~,
\eea
and thus there are no pure gauge nor mixed gauge-gravitational anomalies. There can however be pure gravitational anomalies, corresponding to the non-zero dual spin bordism group 
\bea
\label{eq:Omega9}
\mho^9_{\rm Spin}(pt)= (\ZZ_2)^2~.
\eea
In fact, one part of this group is familiar from the classic work of Alvarez-Gaum{\'e} and Witten \cite{AlvarezGaume:1983ig} -- physically, it corresponds to the fact that a theory of an odd number of Majorana fermions is anomalous in 8d. Since in the present case we have an even number of Majorana fermions, we clearly do not run afoul of this $\ZZ_2$ anomaly. As for the remaining $\ZZ_2$, this is expected to capture an analogous anomaly for an odd number of massless spin-$3 \over 2$ fields, of which we have none.\footnote{A rough explanation for why this anomaly should be related to spin-$3 \over 2$ fields is as follows. Consider compactifying the theory on a closed 8-manifold down to zero dimensions. If there are an odd number of fermionic zero-modes on the 8-manifold, we will encounter the 0d anomaly captured by $\mho^1_{\rm Spin}(pt)= \ZZ_2$. An 8-manifold admits two multiplicative genera, which correspond to the indices of two distinct Dirac operators. One linear combination of these is the usual spin-$\half$ Dirac operator, and the other is the Rarita-Schwinger operator. I thank Kantaro Ohmori for this argument. \label{footnote:anom}} We thus expect the 8d theory to be free of global anomalies.

Finally in $d=9$ we consider the group $\mho^{10}_{\rm Spin}(BE_8)$.  A classic result of Stong \cite{stong1986appendix} gives us 
\bea
\mho^{10}_{\rm Spin}(BE_8) = \mho^{10}_{\rm Spin}(pt) \oplus \tilde \mho^{10}_{\rm Spin}(BE_8) = (\ZZ_2)^3 \oplus (\ZZ_2)^2~.
\eea
In this case we see that there are both potential gauge and gravitational anomalies. As before, one of the pure gravitational anomalies was already discussed in \cite{AlvarezGaume:1983ig} and corresponds to the fact that a theory of an odd number of Majorana fermions is anomalous in 9d. Since we have an even number of fermions, the current setup is not subject to this anomaly. By an argument analogous to that in footnote \ref{footnote:anom}, there is also an anomaly involving an odd number of spin-${3\over 2}$ fields, which causes no trouble here. It remains to fully study the other potential global anomalies.

\subsubsection{Cosmological constant}
Because the $d=6,8,$ and $9$ theories identified above are tachyon-free and non-supersymmetric, one expects them to have a finite and non-zero cosmological constant. 
The one-loop contribution to the cosmological constant may be obtained by simply integrating minus the torus partition function over the fundamental domain $\cF$ \cite{Rohm:1983aq,Polchinski:1985zf}, 
\bea
\Lambda_d = {1 \over 2 (2 \pi)^{d \over 2}}\int_\cF\,{d^2 \tau \over \tau_2^{{d / 2} +1}} (Z_{\rm R}  -Z_{\rm NS})~,
\eea
where we have included the factors of $(2 \pi \tau_2)$ coming from the integral over bosonic zero modes. Though this integral cannot be evaluated analytically, we may evaluate it numerically, giving the following results for the $d=6,8,$ and $9$ theories, 
\bea
\Lambda_6 \approx 0.164~, \hspace{0.5 in} \Lambda_8 \approx 0.079~,\hspace{0.5 in} \Lambda_9 \approx 0.054~,
\eea
in units where $\a' = (2 \pi)^{-1}$. In particular, we see that the one-loop cosmological constant is positive in all cases. 

It is useful to contrast the current situation with that of the $O(16) \times O(16)$ heterotic string. The $O(16) \times O(16)$ string was also tachyon-free and non-supersymmetric, and its one-loop cosmological constant was evaluated in \cite{AlvarezGaume:1986jb}, giving a result $\Lambda_{10} \approx 0.037$ of a similar order of magnitude to the values above. However, in that case there was a crucial conceptual difference in the interpretation of this cosmological constant. A positive cosmological constant generally signals the presence of a dilaton tadpole, which for a massless dilaton gives rise to an instability of the vacuum. In the $O(16) \times O(16)$ heterotic string the dilaton was indeed massless, so despite the fact that the theory was tachyon-free the positive cosmological constant indicated a vacuum instability (see e.g. \cite{Basile:2018irz} for potential stabilization mechanisms). On the other hand, in the case of the lower-dimensional heterotic strings being studied here, the dilaton is actually \textit{massive}. As such a tadpole for it only serves as a small perturbation around the $\Lambda_d >0$ vacua, at least in the weak-coupling regime.\footnote{The intuition is that a tadpole gives rise to modifications of the vacuum of order $ {g \over M^2}$. If the particle with a tadpole is massless, this destroys the vacuum. But if the particle is massive, then as long as we are at sufficiently weak coupling these corrections are small and the vacuum is not appreciably affected.} 

Where exactly is the weak-coupling regime? Since in the lower-dimensional theories the dilaton has a spacelike gradient, see (\ref{eq:dilspacelike}), we conclude that the weakly-coupled region is $X^1 \ll 0$. On the other hand, recall that the lower-dimensional vacuum is really only observed at $X^+\gg 0$, where the integrating out of massive fields is valid. Thus the region which is well-described by a lower-dimensional theory with positive cosmological constant is $X^+ \rightarrow \infty$ and $X^1 \rightarrow -\infty$, i.e. late times and towards the left of the $X^1$-axis.  

We should note that since the lower-dimensional theories have no ``Liouville walls," strings can cross over into the region to the right of the $X^1$-axis. In this region tadpoles may lead to strong-coupling effects which ruin the vacuum, and we should ideally switch to an S-dual description. One could certainly imagine deforming the lower-dimensional theory by adding a Liouville wall, thereby keeping the fields confined to the weakly-coupled region with positive cosmological constant. However, it is unclear what such a wall would uplift to in the full, time-dependent solution.

To summarize, it appears that the lower-dimensional string theories admit perturbatively stable vacua with positive vacuum energy in some appropriate region, and with no moduli. This is at the cost of having a linear dilaton.  It would be interesting to see if the positivity of $\Lambda_d$ persists when higher-loop corrections and non-perturbative effects are taken into account.

\subsection{Tachyon condensation to $d=2$}
\label{sec:hettach2d}
In the cases of $n=0,1$, the ansatz (\ref{eq:Mansatz}) cannot be used to condense all components of the tachyon, since there exist more tachyons than physical spacetime dimensions. We must therefore resort to an alternative mechanism. 

One particular mechanism is to simply condense eight components of the tachyon in the previous manner to reduce to a theory in two dimensions. This automatically gives rise to a stable vacuum since, due to the coupling to the spacelike dilaton gradient, the masses of the remaining components of the tachyon are renormalized to zero. Said in a different way, in two dimensions the boundary conditions at $X_1 \rightarrow -\infty$ (where the linear dilaton theory is weakly coupled) for normalizable states is such that even relevant operators do not give rise to normalizable modes that grow exponentially.\footnote{I thank Simeon Hellerman for very useful correspondence about this point.} 

With this in mind we condense eight components of the tachyon. Integrating out the relevant fields from the partition functions in  (\ref{eq:gaugedZs}), (\ref{eq:gaugedLs}), we obtain precisely the partition functions (\ref{eq:2dnhet1}),(\ref{eq:2dnhet2}) of the known 2d heterotic strings reviewed in Section \ref{sec:2dhetstring}. 
For convenience we recall that the level-matched partition functions for these theories are 
\bea
n&=&0: \hspace{0.2 in} \cZ_{\rm NS} = 24 \hspace{0.46 in} \cZ_{\rm R} = 0
\no\\
n&=&1: \hspace{0.2 in} \cZ_{\rm NS} = 8 \hspace{0.53 in} \cZ_{\rm R} =8
\no\\
n&=&2: \hspace{0.2 in} \cZ_{\rm NS} = 0 \hspace{0.52 in} \cZ_{\rm R} = 12
 \eea
where we have included the $n=2$ case, which had a total of 8 tachyons.

 In more detail, beginning with the ten-dimensional $SO(32)$ theory, we condense eight components of the tachyon, which transforms in the fundamental of $SO(32)$. Doing so breaks the $SO(32)$ and gives rise to a 2d theory with gauge symmetry $SO(24)$. The remaining 24 tachyons become massless due to their coupling to the dilaton, giving rise to the 24 in $\cZ_{\rm NS}$. Since there were no massless fermions in the original  $SO(32)$ theory, there are again none in the resulting 2d theory.
 
 Likewise beginning with the $O(16) \times E_8$ tachyonic theory, we condense eight components of the tachyon, which breaks the gauge group to $O(8) \times E_8$. The remaining 8 tachyons become massless, giving rise to the $8$ contribution in $\cZ_{\rm NS}$. It is easy to see how the $\mathbf{8_s}$ and $\mathbf{8_c}$ fermions descend from the data in (\ref{eq:chiralfermdetails}), upon noting the branching $\mathbf{128} = (\mathbf{8_s},\mathbf{8_s}) \oplus (\mathbf{8_c},\mathbf{8_c})$ under the decomposition $O(16) \supset O(8) \times O(8)  $, and likewise for $\mathbf{128'}$. Finally, in the case of the $O(8) \times O(24)$ theory, we again condense 8 components of the tachyon, breaking completely $O(8)$. In this case we have actually removed all tachyons, and hence we are left with an empty NS sector. There is however a single chiral fermion in the fundamental of $O(24)$. This gives the factor of 12 in $\cZ_{\rm R}$.

Incidentally, note that the technique of reducing down to 2d in this way cannot be applied to the heterotic theories with higher-dimensional vacua, i.e. those with $n>2$. Indeed, it is simple to check that reducing those theories to 2d in the above way gives rise to negative degeneracies in the tentative partition functions. This is in line with the fact that the three 2d heterotic strings discussed above are the only such consistent theories.

Finally, let us address the issue of anomalies in the 2d theories obtained here. Since the field content of the $n=0$ and $n=1$ theories is clearly non-chiral, there are no perturbative anomalies. On the other hand, for the $n=2$ case one has 24 chiral fermions, which on their own do have a chiral anomaly. In order to cancel this, one would expect the presence of a Green-Schwarz term. At first sight this might seem rather mysterious from the higher-dimensional perspective, since as discussed at the end of Section \ref{sec:10dheteroticstrings} the original  $O(8) \times O(24)$ theory had no Green-Schwarz term.  However, in 2d the Green-Schwarz term is simply $S_{\rm GS} \propto \int B$, which can in fact descend from 10d.\footnote{Note by the way that the interpretation of this Green-Schwarz term is rather different from its counterpart in higher dimensions -- in the current case we require that $dH = 1$, and hence that there is a spacetime-filling source for the $B$-field. The role of this source is played by a ``long string"; see \cite{Seiberg:2005nk} for details.}

\section{Type 0 strings}
\label{sec:N11strings}
In this section we study tachyonic Type 0 string theories. We will show that all such tachyonic strings admit two-dimensional stable vacua. For the case of oriented Type 0 strings, this was already shown in \cite{Hellerman:2007fc}. In particular, it was found there that a cascade of tachyon condensations leads to a gradual reduction of dimension, ending eventually with the two-dimensional Type 0 strings introduced in \cite{Takayanagi:2003sm,Douglas:2003up}. In the current section we will extend that result to the eight ${\rm Pin}^-$ Type 0 strings.

Of use to us will be the recent understanding that string theories with different GSO projections can be thought of as differing by the addition of certain topological terms to the worldsheet action \cite{Kaidi:2019pzj,Kaidi:2019tyf}. In the presence of boundaries, these topological terms give rise to anomalous edge modes, which in turn lead to extra structure (e.g. Clifford bundles) carried by the endpoints of open strings. This extra structure modifies the K-theoretic classification of branes in the different GSO-projected theories. 

From this point of view,  the fact that the classification of 2d string theories mimics that of tachyonic 10d theories is in some sense obvious. Beginning with a given 10d worldsheet theory, condensing the tachyon leads to solutions in which certain fields are integrated out. But this process has no effect on the topological term present in the worldsheet action. Thus the lower dimensional string should also admit variants distinguished by such terms. When considering open strings, these terms demand the same anomalous edge modes in 2d as in 10d, and hence the K-theory classification of branes (in the applicable range of worldvolume dimensions) must be the same. We will give slightly more detail on this below.

To recapitulate, for many known two-dimensional closed superstrings, we will identify a ten-dimensional tachyonic string which admits a dimension-changing dynamical transition to them. It should be noted that this hierarchy can be continued upwards indefinitely, i.e. the ten-dimensional theory can itself be understood as descending from condensation of a supercritical theory, and so on. The only special thing about ten-dimensions is that a linear dilaton is not needed for worldsheet anomaly cancellation.

\subsection{Tachyon condensation}

We begin by reviewing the general framework of tachyon condensation for $\cN=(1,1)$ superstrings, which differs slightly from that in the heterotic case.

 For the known tachyonic $\cN=(1,1)$ superstrings, there is always a single tachyon. Condensation of the tachyon is expected to give a $(1,1)$ superpotential,
\bea
W = \cT(X)~,
\eea
which gives rise to a scalar potential 
\bea
V = {\a' \over 16 \pi} g^{\m \n} \p_\m \cT(X) \p_\n \cT(X) - {i \a' \over 4 \pi} \p_\m \p_\n \cT(X) \tilde{\psi}^\m \psi^\n~.
\eea
Unlike in (\ref{eq:hetscalarpot}), this scalar potential depends only on the derivatives of the tachyon. Hence we will have to take the tachyon profile to be quadratic, as opposed to linear, in the fields that we want to give a mass to. Taking the dilaton to be as in (\ref{eq:higherdimdil}), the simplest tachyon profile (subject to some additional discrete global symmetry constraints) is of the form \cite{Hellerman:2006ff}
\bea
\label{eq:N11tachprof}
\cT(X) = {2 m \over \a'} \,e^{\b X_+} \sum_{i=2}^{r+2}  X^i X^{r+i}
\eea
for some $r$. This profile then gives rise to exponentially growing masses for the $2r$ coordinates $X^i$ with $i=2, \dots, 2r+2$, as well as their fermionic superpartners $\psi^i$. Consequently as $X^+ \rightarrow \infty$ we obtain a theory in $d=10-2r$ dimensions, localized at $X^2 = \dots = X^{2r+2}=0$.

In this way we may condense ten-dimensional $\cN=(1,1)$ theories to any even number of dimensions (for the case of an odd number of dimensions, see \cite{Hellerman:2007fc}). However, by explicitly integrating out the appropriate fields in the ten-dimensional partition functions given below, it is simple to check that a new tachyon appears if we condense to any dimension $d>2$. One thus obtains a cascade of tachyon condensations, the stable endpoint of which is a two-dimensional vacuum. We now show that these are precisely the known two-dimensional strings. 

 \subsection{Oriented Type 0 strings}

\subsubsection{$d=10$}
We begin by reviewing the known case of oriented Type 0 strings. 
These strings have the same worldsheet content as the Type II strings, namely ten left- and right-moving bosonic fields $X^\m$, together with their superpartners $\psi^\m$ and $\tilde \psi^\m$. The defining feature of Type 0 strings is the fact that left- and right-moving worldsheet fermions are taken to have the same spin structures. 
Because of this identification of spin structures, the torus partition function is 
\bea
\label{eq:Type0partfunct}
Z = \half |\eta|^{-16}\left(  |Z^0_0|^{16} + |Z^0_1|^{16}+ |Z^1_0|^{16} \pm |Z^1_1|^{16}\right)
\eea
in the notation of (\ref{eq:Zabdef}). Above we have included a possible choice of sign in front of the final term. This sign has no effect on the partition function since $Z^1_1 = 0$, and the level-matched result is
\bea
\cZ= (q \bar q)^{-\half} + 192 + 1296 (q \bar q)^{\half} + 49152\, q \bar q +\dots 
\eea
in both cases. We see that both theories have a single tachyon. There are no spacetime fermions in these theories, so the 192 massless degrees of freedom can decomposed into a graviton (35), a $B$-field (28), a dilaton (1), and a remaining 128 massless bosonic degrees of freedom which end up being RR form fields. 

The difference between the two signs is in how the 128 degrees of freedom are organized into separate RR form fields. As was discussed in \cite{Kaidi:2019pzj,Kaidi:2019tyf}, this may be understood by interpreting the sign as arising due to the presence of an extra topological term in the worldsheet action. Concretely, given a worldsheet $\Sigma$ and a spin structure $\sigma$, we may add to the action a factor of the Arf invariant,\footnote{We consider only the cases of zero or one copy of $\Arf$ since it is a mod 2 invariant.}
\bea
\label{eq:Arfinvws}
S = S_0 + i \pi n\, \Arf(\Sigma, \sigma)~,\hspace{0.5 in} n=0,1
\eea
where $S_0$ is the standard worldsheet action without topological term. Concretely, the Arf invariant is defined such that \cite{arf1941untersuchungen}
\bea
\Arf(T^2, \sigma_{\rm NS NS}) = \Arf(T^2, \sigma_{\rm NS R})=\Arf(T^2, \sigma_{\rm R NS})=0 ~, \hspace{0.3 in}\Arf(T^2, \sigma_{\rm R R}) = 1~,
\eea
and hence the contribution of this topological term to the torus partition function, namely $(-1)^{n \,\Arf(T^2, \sigma)}$, reproduces the sign in (\ref{eq:Type0partfunct}).
In the presence of a boundary, the $\Arf$ invariant leads to Majorana edge modes, which modify the K-theory classification of D-branes and RR fields.\footnote{This mechanism was  anticipated in \cite{Witten:1998cd,WittenStanford,WittenSCGP}.}

The upshot is that for the plus sign, the 128 degrees of freedom are organized into two 1-forms and two 3-forms, while for the minus sign they are organized into two scalars, two 2-forms, and a non-self-dual 4-form. This field content is exactly twice that of the usual Type IIA and IIB theories, and for that reason these theories are usual referred to as Type 0A and 0B. The spectrum of branes in a given Type 0 theory is twice that of its Type II counterpart. 

\subsubsection{$d=2$}
\label{sec:2dType0}
Two-dimensional analogs of the Type 0 theories have a long history, see e.g. \cite{Takayanagi:2003sm,Douglas:2003up,Seiberg:2005nk}. In two dimensions, the Type 0B theory has as its physical operator spectrum one propagating massless bosonic field in the $\rm{NSNS}$ sector (often referred to as a ``tachyon", even though it is massless in two dimensions) as well as a pair of compact chiral bosons in the $\rm RR$ sector. As for the Type 0A theory, there is again a single scalar in the $\rm{NSNS}$ sector, as well as a pair of 1-form fields (fixed at zero momentum). 

In  \cite{Hellerman:2007fc}, it was shown that the ten-dimensional Type 0A/B theory condenses to the two-dimensional Type 0A/B theory upon choosing the profile (\ref{eq:N11tachprof}) with $r=4$. The massless  $\rm{NSNS}$ sector scalar is the descendant of the tachyon in higher dimensions. The fact that the spectra of RR fields match with their higher dimensional counterparts (for $p$-forms with $p<2$) follows from the fact that the worldsheet factors of the Arf invariant in (\ref{eq:Arfinvws}) are left unchanged by the integrating out of fields, and lead to the same spectra of boundary Majorana fermions.

\label{sec:2dType0}
\subsection{Unoriented Type 0 Strings}
\subsubsection{$d=10$}
We now move on to the new case of unoriented Type 0 theories. In \cite{Kaidi:2019pzj,Kaidi:2019tyf} unoriented Type 0 theories were classified, again using invertible phases.\footnote{For previous works towards such a classification, see e.g. \cite{Sagnotti:1995ga,Sagnotti:1996qj,Bergman:1997rf,Bergman:1999km}.} There are two broad classes of unoriented Type 0 strings -- those with $\rm{Pin}^-$ structure on the worldsheet, and those with  $\rm{Pin}^+$ structure.  Recall that the difference between $\rm{Pin}^\pm$ can be captured by the squaring properties of the orientation reversal operation $\rm R$ being gauged, i.e.
\bea
\rm{Pin}^+: \hspace{0.2 in} R^2 = 1~, \hspace{0.6 in}\rm{Pin}^-: \hspace{0.2 in} R^2 = (-1)^{\f}~,
\eea
with $(-1)^{\f}:=(-1)^{\fl + \fr}$ the total worldsheet fermion number. In the standard conventions, worldsheet parity $\Omega$ acts on fermions as 
\bea
\Omega \,\psi(t, \sigma)\, \Omega^{-1} = - \tilde{\psi}(t, 2 \pi - \sigma)~, \hspace{0.5 in} \Omega\, \tilde{\psi}(t, \sigma) \,\Omega^{-1} = \psi(t, 2\pi - \sigma)~,
\eea
and hence $\Omega^2 = (-1)^{\f}$. Thus gauging $\Omega$ alone gives rise to worldsheets with $\rm{Pin}^-$ structure. Unlike for Type II theories, $\Omega$ is a symmetry of both Type 0A/B, and hence can be gauged in both to give two separate $\rm{Pin}^-$  theories. 

One can also consider gauging $\Omega$ twisted by some additional discrete global symmetries. In particular, the Type 0 theories have a $\ZZ_2 \times \ZZ_2$ symmetry generated by left-moving spacetime and worldsheet fermion parities, $(-1)^{\FL}$ and $(-1)^{\fl}$. We may then consider alternative orientifoldings by $\Omega_\F := \Omega (-1)^{\FL}$ and $\Omega_\f := \Omega (-1)^{\fl}$. Note that $(\Omega_{\F})^2 = (-1)^{\f}$ once again, and thus gauging it in the Type 0A/B theories again gives two $\Pin^-$ theories. In fact, each of the four  $\Pin^-$ theories mentioned thus far admits two variants, differing by the action of $\Omega$ on Chan-Paton factors; this is analogous to the difference between ${\rm{O9}}^-$ and ${\rm{O9}}^+$ orientifolds of Type IIB. So in total there are actually eight $\Pin^-$ Type 0 strings. 

In \cite{Kaidi:2019pzj,Kaidi:2019tyf}, these eight theories were interpreted as arising from the addition of a topological term to the worldsheet action. In particular, given a worldsheet $\Sigma$ with $\Pin^-$ structure $\sigma$, one can consider adding to the action $n$ copies of the so-called ABK invariant \cite{brown1972generalizations}, much as for the Arf invariant in the oriented case. The ABK invariant is a mod 8 invariant, and we have the following identifications, 
\bea
\label{eq:ndef}
n&=&0,4: \quad ({\rm{0B}}, \Omega) \hspace{0.7 in}n=1,5: \quad (\rm{0A}, \Omega)
\no\\
n&=&2,6: \quad ({\rm{0B}}, \Omega_{\F}) \hspace{0.635 in}n=3,7: \quad (\rm{0A}, \Omega_{\F})
\eea
with the first element in parenthesis denoting the starting theory, and the second element denoting the operator being gauged. The difference between theories differing by $\Delta n = 4$ is the action of the parity operators on Chan-Paton factors. The massless and tachyonic spectra of all 8 theories was determined in  \cite{Kaidi:2019pzj,Kaidi:2019tyf} -- for our purposes, we simply note that all theories have a single tachyon, and the spectrum of branes given in Table \ref{branetable}.

\begin{table}[thp]
\begin{center}
\begin{tabular}{c|ccccccccccc}
\hline
$n\,\backslash\, p$   & $-1$& $0$ &   $1$ &    $2$& $3$&   $4$ & $5$ &$6$ & $7$& $8$ & $9$
\\\hline
0 & $2\ZZ_2$ & $2\ZZ_2$ & $2\ZZ$ &0  &0 & 0 &$2\ZZ$&0&$2\ZZ_2$&$2\ZZ_2$&$2\ZZ$ 
\\
1 &  $\ZZ_2$ & $ \ZZ {\oplus} \ZZ_2$ &$\ZZ_2$  &$\ZZ$  & $0$  &$ \ZZ $&$0$ &$\ZZ{\oplus} \ZZ_2$ &$\ZZ_2$&$\ZZ{\oplus} \ZZ_2$&$\ZZ_2$ 
\\
2 & $2\ZZ$ & $0$ & $\ZZ_2$ &$\ZZ_2$  &$2\ZZ $ & $0$ &$ \ZZ_2 $&$\ZZ_2$&$2\ZZ$&$0$&$\ZZ_2$ 
\\
3 & $0$ & $\ZZ$ & $0$  &$\ZZ{\oplus}\ZZ_2$ & $\ZZ_2$ &$\ZZ{\oplus} \ZZ_2$&$ \ZZ_2$&$\ZZ$&$0$&$\ZZ$&$0$
\\
4& 0& 0 & $2\ZZ$ &0  &$2\ZZ_2$ & $2\ZZ_2$ &$2\ZZ$&0&0&0&$2\ZZ$ 
\\
5 &$0$ & $\ZZ$ & $0$  &$\ZZ{\oplus}\ZZ_2$ & $\ZZ_2$ &$\ZZ{\oplus} \ZZ_2$&$ \ZZ_2$&$\ZZ$&$0$&$\ZZ$&$0$
\\
6 & $2\ZZ$ & $0$ & $\ZZ_2$ &$\ZZ_2$  &$2\ZZ $ & $0$ &$ \ZZ_2 $&$\ZZ_2$&$2\ZZ$&$0$&$\ZZ_2$ 
\\
7 &  $\ZZ_2$ & $ \ZZ {\oplus} \ZZ_2$ &$\ZZ_2$  &$\ZZ$  & $0$  &$ \ZZ $&$0$ &$\ZZ{\oplus} \ZZ_2$ &$\ZZ_2$&$\ZZ{\oplus} \ZZ_2$&$\ZZ_2$ 
\\\hline
\end{tabular}
\end{center}
\caption{${\rm D}p$-brane content of the eight $\rm{Pin}^-$ Type 0 theories. }
\label{branetable}
\end{table}%

Before proceeding to two dimensions, we address the remaining parity operator $\Omega_\f$. It is clear that
\bea
(\Omega_\f)^2 = \Omega (-1)^{\fl}\Omega (-1)^{\fl} = \Omega (-1)^{\fl + \fr} \Omega = (-1)^{\f} \Omega^2 = 1~,
\eea
so this gives a theory with $\Pin^+$ structure. In \cite{Kaidi:2019tyf} this case was shown to be tachyon-free, so it will not be relevant for our discussion.
 
 \subsubsection{$d=2$}
 \label{sec:2dPin-}
The $\Pin^-$ theories discussed above condense to two-dimensional unoriented Type 0 theories. The latter were classified in \cite{Bergman:2003yp,Gomis:2005ce}. As expected, they are obtained by starting with the 2d Type 0 theories of Section \ref{sec:2dType0} and gauging an appropriate parity operation.

In  \cite{Bergman:2003yp,Gomis:2005ce}, eight 2d $\Pin^-$ strings were identified, again interpretable as orientifolds of 2d Type 0A/B by $\Omega$ or $\Omega_{\F}$, with some two-fold choice of action on Chan-Paton factors. We may again label the theories by $n \in \{0, \dots, 7\}$, in accordance with the assignments in (\ref{eq:ndef}). In \cite{Bergman:2003yp} the spectra of these theories were found to be as follows. In the $n=0$ theory there is a single propagating massless scalar field from the $\rm NSNS$ sector, as well as two 2-forms (of fixed momentum). The objects charged under these 2-forms are two distinct types of D1-branes. In the $n=1$ theory, there is again a single propagating massless scalar field, as well as a single 1-form. The object charged under this is a single type of D0-brane. In the $n=2$ theory, there are two massless scalar fields, one of which is compact. There are ${\rm D}(-1)$-branes charged under the compact scalar. Finally, in the $n=3$ case there is again a single propagating massless scalar field, as well as a single 1-form. The object charged under this is a single type of D0-brane. The (non-torsion) spectrum is periodic mod 4, so this completes the list.
 
 In all cases, the massless scalar field can be interpreted as the descendant of the higher-dimensional tachyon upon condensation. The spectra of branes is seen to match precisely with the non-torsion elements listed in Table \ref{branetable}, in the appropriate range of worldvolume dimensions. This again must be the case, since the worldsheet factors of $\rm ABK$ -- and hence the corresponding K-theory classification --  remain unchanged upon condensation of the tachyon.

\section*{Acknowledgements}
I would like to thank Luis Alvarez-Gaum{\'e}, Oren Bergman, Simeon Hellerman, Per Kraus, and Kantaro Ohmori for very useful discussions at various stages of this project. I am especially grateful to Simeon Hellerman for information regarding two-dimensional strings.
\bibliographystyle{JHEP}
\bibliography{bib}

\providecommand{\href}[2]{#2}\begingroup\raggedright\begin{thebibliography}{10}

\bibitem{Bergman:1997rf}
O.~Bergman and M.~R. Gaberdiel, \emph{{A Nonsupersymmetric open string theory
  and S duality}},
  \href{http://dx.doi.org/10.1016/S0550-3213(97)00309-X}{\emph{Nucl. Phys. B}
  {\bf 499} (1997) 183--204}, [\href{http://arxiv.org/abs/hep-th/9701137}{{\tt
  hep-th/9701137}}].

\bibitem{Bergman:1999km}
O.~Bergman and M.~R. Gaberdiel, \emph{{Dualities of type 0 strings}},
  \href{http://dx.doi.org/10.1088/1126-6708/1999/07/022}{\emph{JHEP} {\bf 07}
  (1999) 022}, [\href{http://arxiv.org/abs/hep-th/9906055}{{\tt
  hep-th/9906055}}].

\bibitem{Kaidi:2019pzj}
J.~Kaidi, J.~Parra-Martinez and Y.~Tachikawa, \emph{{Classification of String
  Theories via Topological Phases}},
  \href{http://dx.doi.org/10.1103/PhysRevLett.124.121601}{\emph{Phys. Rev.
  Lett.} {\bf 124} (2020) 121601}, [\href{http://arxiv.org/abs/1908.04805}{{\tt
  1908.04805}}].

\bibitem{Kaidi:2019tyf}
J.~Kaidi, J.~Parra-Martinez and Y.~Tachikawa, \emph{{Topological
  Superconductors on Superstring Worldsheets}},
  \href{http://dx.doi.org/10.21468/SciPostPhys.9.1.010}{\emph{SciPost Phys.}
  {\bf 9} (2020) 10}, [\href{http://arxiv.org/abs/1911.11780}{{\tt
  1911.11780}}].

\bibitem{AlvarezGaume:1986jb}
L.~Alvarez-Gaume, P.~H. Ginsparg, G.~W. Moore and C.~Vafa, \emph{{An O(16)
  $\times$ O(16) Heterotic String}},
  \href{http://dx.doi.org/10.1016/0370-2693(86)91524-8}{\emph{Phys. Lett. B}
  {\bf 171} (1986) 155--162}.

\bibitem{Dixon:1986iz}
L.~J. Dixon and J.~A. Harvey, \emph{{String Theories in Ten-Dimensions Without
  Space-Time Supersymmetry}},
  \href{http://dx.doi.org/10.1016/0550-3213(86)90619-X}{\emph{Nucl. Phys. B}
  {\bf 274} (1986) 93--105}.

\bibitem{Kawai:1986vd}
H.~Kawai, D.~Lewellen and S.~Tye, \emph{{Classification of Closed Fermionic
  String Models}},
  \href{http://dx.doi.org/10.1103/PhysRevD.34.3794}{\emph{Phys. Rev. D} {\bf
  34} (1986) 3794}.

\bibitem{Itoyama:1986ei}
H.~Itoyama and T.~Taylor, \emph{{Supersymmetry Restoration in the Compactified
  O(16)$\times$O(16)-prime Heterotic String Theory}},
  \href{http://dx.doi.org/10.1016/0370-2693(87)90267-X}{\emph{Phys. Lett. B}
  {\bf 186} (1987) 129--133}.

\bibitem{Ginsparg:1986wr}
P.~H. Ginsparg and C.~Vafa, \emph{{Toroidal Compactification of
  Nonsupersymmetric Heterotic Strings}},
  \href{http://dx.doi.org/10.1016/0550-3213(87)90387-7}{\emph{Nucl. Phys. B}
  {\bf 289} (1987) 414}.

\bibitem{Blum:1997cs}
J.~D. Blum and K.~R. Dienes, \emph{{Duality without supersymmetry: The Case of
  the SO(16) $\times$SO(16) string}},
  \href{http://dx.doi.org/10.1016/S0370-2693(97)01172-6}{\emph{Phys. Lett. B}
  {\bf 414} (1997) 260--268}, [\href{http://arxiv.org/abs/hep-th/9707148}{{\tt
  hep-th/9707148}}].

\bibitem{Blum:1997gw}
J.~D. Blum and K.~R. Dienes, \emph{{Strong / weak coupling duality relations
  for nonsupersymmetric string theories}},
  \href{http://dx.doi.org/10.1016/S0550-3213(97)00803-1}{\emph{Nucl. Phys. B}
  {\bf 516} (1998) 83--159}, [\href{http://arxiv.org/abs/hep-th/9707160}{{\tt
  hep-th/9707160}}].

\bibitem{Suyama:2001bn}
T.~Suyama, \emph{{Closed string tachyons in nonsupersymmetric heterotic
  theories}},
  \href{http://dx.doi.org/10.1088/1126-6708/2001/08/037}{\emph{JHEP} {\bf 08}
  (2001) 037}, [\href{http://arxiv.org/abs/hep-th/0106079}{{\tt
  hep-th/0106079}}].

\bibitem{Hellerman:2007fc}
S.~Hellerman and I.~Swanson, \emph{{Charting the landscape of supercritical
  string theory}},
  \href{http://dx.doi.org/10.1103/PhysRevLett.99.171601}{\emph{Phys. Rev.
  Lett.} {\bf 99} (2007) 171601}, [\href{http://arxiv.org/abs/0705.0980}{{\tt
  0705.0980}}].

\bibitem{Faraggi:2019fap}
A.~E. Faraggi, \emph{{String Phenomenology From a Worldsheet Perspective}},
  \href{http://dx.doi.org/10.1140/epjc/s10052-019-7222-5}{\emph{Eur. Phys. J.
  C} {\bf 79} (2019) 703}, [\href{http://arxiv.org/abs/1906.09448}{{\tt
  1906.09448}}].

\bibitem{Faraggi:2019drl}
A.~E. Faraggi, V.~G. Matyas and B.~Percival, \emph{{Stable Three Generation
  Standard--like Model From a Tachyonic Ten Dimensional Heterotic--String
  Vacuum}}, \href{http://dx.doi.org/10.1140/epjc/s10052-020-7894-x}{\emph{Eur.
  Phys. J. C} {\bf 80} (2020) 337},
  [\href{http://arxiv.org/abs/1912.00061}{{\tt 1912.00061}}].

\bibitem{Faraggi:2020wej}
A.~E. Faraggi, V.~G. Matyas and B.~Percival, \emph{{Towards the Classification
  of Tachyon-Free Models From Tachyonic Ten-Dimensional Heterotic String
  Vacua}},  \href{http://arxiv.org/abs/2006.11340}{{\tt 2006.11340}}.

\bibitem{Sugimoto:1999tx}
S.~Sugimoto, \emph{{Anomaly cancellations in type I D-9 - anti-D-9 system and
  the USp(32) string theory}},
  \href{http://dx.doi.org/10.1143/PTP.102.685}{\emph{Prog. Theor. Phys.} {\bf
  102} (1999) 685--699}, [\href{http://arxiv.org/abs/hep-th/9905159}{{\tt
  hep-th/9905159}}].

\bibitem{Hellerman:2004qa}
S.~Hellerman and X.~Liu, \emph{{Dynamical dimension change in supercritical
  string theory}},  \href{http://arxiv.org/abs/hep-th/0409071}{{\tt
  hep-th/0409071}}.

\bibitem{Hellerman:2006nx}
S.~Hellerman and I.~Swanson, \emph{{Cosmological solutions of supercritical
  string theory}},
  \href{http://dx.doi.org/10.1103/PhysRevD.77.126011}{\emph{Phys. Rev. D} {\bf
  77} (2008) 126011}, [\href{http://arxiv.org/abs/hep-th/0611317}{{\tt
  hep-th/0611317}}].

\bibitem{Hellerman:2006ff}
S.~Hellerman and I.~Swanson, \emph{{Dimension-changing exact solutions of
  string theory}},
  \href{http://dx.doi.org/10.1088/1126-6708/2007/09/096}{\emph{JHEP} {\bf 09}
  (2007) 096}, [\href{http://arxiv.org/abs/hep-th/0612051}{{\tt
  hep-th/0612051}}].

\bibitem{Hellerman:2006hf}
S.~Hellerman and I.~Swanson, \emph{{Cosmological unification of string
  theories}},
  \href{http://dx.doi.org/10.1088/1126-6708/2008/07/022}{\emph{JHEP} {\bf 07}
  (2008) 022}, [\href{http://arxiv.org/abs/hep-th/0612116}{{\tt
  hep-th/0612116}}].

\bibitem{Hellerman:2007zz}
S.~Hellerman and I.~Swanson, \emph{{A Stable vacuum of the tachyonic E(8)
  string}},  \href{http://arxiv.org/abs/0710.1628}{{\tt 0710.1628}}.

\bibitem{McGuigan:1991qp}
M.~D. McGuigan, C.~R. Nappi and S.~A. Yost, \emph{{Charged black holes in
  two-dimensional string theory}},
  \href{http://dx.doi.org/10.1016/0550-3213(92)90039-E}{\emph{Nucl. Phys. B}
  {\bf 375} (1992) 421--450}, [\href{http://arxiv.org/abs/hep-th/9111038}{{\tt
  hep-th/9111038}}].

\bibitem{Giveon:2004zz}
A.~Giveon and A.~Sever, \emph{{Strings in a 2-d extremal black hole}},
  \href{http://dx.doi.org/10.1088/1126-6708/2005/02/065}{\emph{JHEP} {\bf 02}
  (2005) 065}, [\href{http://arxiv.org/abs/hep-th/0412294}{{\tt
  hep-th/0412294}}].

\bibitem{Davis:2005qe}
J.~L. Davis, F.~Larsen and N.~Seiberg, \emph{{Heterotic strings in two
  dimensions and new stringy phase transitions}},
  \href{http://dx.doi.org/10.1088/1126-6708/2005/08/035}{\emph{JHEP} {\bf 08}
  (2005) 035}, [\href{http://arxiv.org/abs/hep-th/0505081}{{\tt
  hep-th/0505081}}].

\bibitem{Seiberg:2005nk}
N.~Seiberg, \emph{{Long strings, anomaly cancellation, phase transitions,
  T-duality and locality in the 2-D heterotic string}},
  \href{http://dx.doi.org/10.1088/1126-6708/2006/01/057}{\emph{JHEP} {\bf 01}
  (2006) 057}, [\href{http://arxiv.org/abs/hep-th/0511220}{{\tt
  hep-th/0511220}}].

\bibitem{Davis:2005qi}
J.~L. Davis, \emph{{The Moduli space and phase structure of heterotic strings
  in two dimensions}},
  \href{http://dx.doi.org/10.1103/PhysRevD.74.026004}{\emph{Phys. Rev. D} {\bf
  74} (2006) 026004}, [\href{http://arxiv.org/abs/hep-th/0511298}{{\tt
  hep-th/0511298}}].

\bibitem{Takayanagi:2003sm}
T.~Takayanagi and N.~Toumbas, \emph{{A Matrix model dual of type 0B string
  theory in two-dimensions}},
  \href{http://dx.doi.org/10.1088/1126-6708/2003/07/064}{\emph{JHEP} {\bf 07}
  (2003) 064}, [\href{http://arxiv.org/abs/hep-th/0307083}{{\tt
  hep-th/0307083}}].

\bibitem{Douglas:2003up}
M.~R. Douglas, I.~R. Klebanov, D.~Kutasov, J.~M. Maldacena, E.~J. Martinec and
  N.~Seiberg, \emph{{A New hat for the c=1 matrix model}},
  \href{http://arxiv.org/abs/hep-th/0307195}{{\tt hep-th/0307195}}.

\bibitem{Bergman:2003yp}
O.~Bergman and S.~Hirano, \emph{{The Cap in the hat: Unoriented 2-D strings and
  matrix (vector) models}},
  \href{http://dx.doi.org/10.1088/1126-6708/2004/01/043}{\emph{JHEP} {\bf 01}
  (2004) 043}, [\href{http://arxiv.org/abs/hep-th/0311068}{{\tt
  hep-th/0311068}}].

\bibitem{Gomis:2005ce}
J.~Gomis, \emph{{Anomaly cancellation in noncritical string theory}},
  \href{http://dx.doi.org/10.1088/1126-6708/2005/10/095}{\emph{JHEP} {\bf 10}
  (2005) 095}, [\href{http://arxiv.org/abs/hep-th/0508132}{{\tt
  hep-th/0508132}}].

\bibitem{McGreevy:2003kb}
J.~McGreevy and H.~L. Verlinde, \emph{{Strings from tachyons: The c=1 matrix
  reloaded}},
  \href{http://dx.doi.org/10.1088/1126-6708/2003/12/054}{\emph{JHEP} {\bf 12}
  (2003) 054}, [\href{http://arxiv.org/abs/hep-th/0304224}{{\tt
  hep-th/0304224}}].

\bibitem{Gomis:2003vi}
J.~Gomis and A.~Kapustin, \emph{{Two-dimensional unoriented strings and matrix
  models}}, \href{http://dx.doi.org/10.1088/1126-6708/2004/06/002}{\emph{JHEP}
  {\bf 06} (2004) 002}, [\href{http://arxiv.org/abs/hep-th/0310195}{{\tt
  hep-th/0310195}}].

\bibitem{Takayanagi:2004ge}
T.~Takayanagi, \emph{{Comments on 2-D type IIA string and matrix model}},
  \href{http://dx.doi.org/10.1088/1126-6708/2004/11/030}{\emph{JHEP} {\bf 11}
  (2004) 030}, [\href{http://arxiv.org/abs/hep-th/0408086}{{\tt
  hep-th/0408086}}].

\bibitem{Seiberg:2005bx}
N.~Seiberg, \emph{{Observations on the moduli space of two dimensional string
  theory}}, \href{http://dx.doi.org/10.1088/1126-6708/2005/03/010}{\emph{JHEP}
  {\bf 03} (2005) 010}, [\href{http://arxiv.org/abs/hep-th/0502156}{{\tt
  hep-th/0502156}}].

\bibitem{Ita:2005ne}
H.~Ita, H.~Nieder and Y.~Oz, \emph{{On type II strings in two dimensions}},
  \href{http://dx.doi.org/10.1088/1126-6708/2005/06/055}{\emph{JHEP} {\bf 06}
  (2005) 055}, [\href{http://arxiv.org/abs/hep-th/0502187}{{\tt
  hep-th/0502187}}].

\bibitem{Lerche:1986he}
W.~Lerche and D.~Lust, \emph{{Covariant Heterotic Strings and Odd Selfdual
  Lattices}}, \href{http://dx.doi.org/10.1016/0370-2693(87)90069-4}{\emph{Phys.
  Lett. B} {\bf 187} (1987) 45--50}.

\bibitem{Lerche:1986ae}
W.~Lerche, D.~Lust and A.~Schellekens, \emph{{Ten-dimensional Heterotic Strings
  From Niemeier Lattices}},
  \href{http://dx.doi.org/10.1016/0370-2693(86)91257-8}{\emph{Phys. Lett. B}
  {\bf 181} (1986) 71}.

\bibitem{Lust:1989tj}
D.~Lust and S.~Theisen, \emph{{Lectures on string theory}}, vol.~346.
\newblock 1989,
  \href{http://dx.doi.org/10.1007/BFb0113507}{10.1007/BFb0113507}.

\bibitem{chamseddine1992study}
A.~H. Chamseddine, \emph{A study of non-critical strings in arbitrary
  dimensions}, {\emph{Nuclear Physics B} {\bf 368} (1992) 98--120}.

\bibitem{Erler:1993zy}
J.~Erler, \emph{{Anomaly cancellation in six-dimensions}},
  \href{http://dx.doi.org/10.1063/1.530885}{\emph{J. Math. Phys.} {\bf 35}
  (1994) 1819--1833}, [\href{http://arxiv.org/abs/hep-th/9304104}{{\tt
  hep-th/9304104}}].

\bibitem{Ohmori:2014kda}
K.~Ohmori, H.~Shimizu, Y.~Tachikawa and K.~Yonekura, \emph{{Anomaly polynomial
  of general 6d SCFTs}},
  \href{http://dx.doi.org/10.1093/ptep/ptu140}{\emph{PTEP} {\bf 2014} (2014)
  103B07}, [\href{http://arxiv.org/abs/1408.5572}{{\tt 1408.5572}}].

\bibitem{Sagnotti:1992qw}
A.~Sagnotti, \emph{{A Note on the Green-Schwarz mechanism in open string
  theories}}, \href{http://dx.doi.org/10.1016/0370-2693(92)90682-T}{\emph{Phys.
  Lett. B} {\bf 294} (1992) 196--203},
  [\href{http://arxiv.org/abs/hep-th/9210127}{{\tt hep-th/9210127}}].

\bibitem{edwards1991spin}
S.~R. Edwards et~al., \emph{On the spin bordism of {$B (E_8 \times E_8)$}},
  {\emph{Illinois journal of mathematics} {\bf 35} (1991) 683--689}.

\bibitem{Lee:2020ojw}
Y.~Lee, K.~Ohmori and Y.~Tachikawa, \emph{{Revisiting Wess-Zumino-Witten
  terms}},  \href{http://arxiv.org/abs/2009.00033}{{\tt 2009.00033}}.

\bibitem{AlvarezGaume:1983ig}
L.~Alvarez-Gaume and E.~Witten, \emph{{Gravitational Anomalies}},
  \href{http://dx.doi.org/10.1016/0550-3213(84)90066-X}{\emph{Nucl. Phys. B}
  {\bf 234} (1984) 269}.

\bibitem{stong1986appendix}
R.~Stong, \emph{Appendix: calculation of {$\Omega^{Spin}_{11}(K (Z, 4))$}},  in
  \emph{Workshop on unified string theories (Santa Barbara, Calif., 1985)},
  vol.~430437, 1986.

\bibitem{Rohm:1983aq}
R.~Rohm, \emph{{Spontaneous Supersymmetry Breaking in Supersymmetric String
  Theories}}, \href{http://dx.doi.org/10.1016/0550-3213(84)90007-5}{\emph{Nucl.
  Phys. B} {\bf 237} (1984) 553--572}.

\bibitem{Polchinski:1985zf}
J.~Polchinski, \emph{{Evaluation of the One Loop String Path Integral}},
  \href{http://dx.doi.org/10.1007/BF01210791}{\emph{Commun. Math. Phys.} {\bf
  104} (1986) 37}.

\bibitem{Basile:2018irz}
I.~Basile, J.~Mourad and A.~Sagnotti, \emph{{On Classical Stability with Broken
  Supersymmetry}}, \href{http://dx.doi.org/10.1007/JHEP01(2019)174}{\emph{JHEP}
  {\bf 01} (2019) 174}, [\href{http://arxiv.org/abs/1811.11448}{{\tt
  1811.11448}}].

\bibitem{arf1941untersuchungen}
C.~Arf, \emph{Untersuchungen {\"u}ber quadratische formen in k{\"o}rpern der
  charakteristik 2.(teil i.).}, {\emph{Journal f{\"u}r die reine und angewandte
  Mathematik} {\bf 1941} (1941) 148--167}.

\bibitem{Witten:1998cd}
E.~Witten, \emph{{D-branes and K theory}},
  \href{http://dx.doi.org/10.1088/1126-6708/1998/12/019}{\emph{JHEP} {\bf 12}
  (1998) 019}, [\href{http://arxiv.org/abs/hep-th/9810188}{{\tt
  hep-th/9810188}}].

\bibitem{WittenStanford}
E.~Witten, \emph{Nonsupersymmetric d-branes and the kitaev fermion chain}, .

\bibitem{WittenSCGP}
E.~Witten, \emph{Anomalies and nonsupersymmetric d-branes}, .

\bibitem{Sagnotti:1995ga}
A.~Sagnotti, \emph{{Some properties of open string theories}},  in
  \emph{{International Workshop on Supersymmetry and Unification of Fundamental
  Interactions (SUSY 95)}}, pp.~473--484, 9, 1995.
\newblock \href{http://arxiv.org/abs/hep-th/9509080}{{\tt hep-th/9509080}}.

\bibitem{Sagnotti:1996qj}
A.~Sagnotti, \emph{{Surprises in open string perturbation theory}},
  \href{http://dx.doi.org/10.1016/S0920-5632(97)00344-7}{\emph{Nucl. Phys. B
  Proc. Suppl.} {\bf 56} (1997) 332--343},
  [\href{http://arxiv.org/abs/hep-th/9702093}{{\tt hep-th/9702093}}].

\bibitem{brown1972generalizations}
E.~H. Brown, \emph{Generalizations of the kervaire invariant}, {\emph{Annals of
  Mathematics} (1972) 368--383}.

\end{thebibliography}\endgroup
\end{document}